\documentclass[aps,pre,showpacs,lineno,groupedaddress]{revtex4}
\usepackage[dvips]{graphicx} \usepackage{float}
\usepackage{dcolumn} % needed for some tables
\usepackage{bm} % for math
\usepackage{amssymb} % for math

\begin{document}
\title{\textit{In silico} network topology-based prediction of gene
  essentiality}

\author{Jo\~ao Paulo M\"uller da Silva}
\affiliation{Department of Physics and Biophysics, Institute of Biosciences, S\~ao Paulo State University, UNESP, 18618-000, Botucatu, SP, Brazil} 

\author{Marcio Luis Acencio}
\affiliation{Department of Physics and Biophysics, Institute of Biosciences, S\~ao Paulo State University, UNESP, 18618-000, Botucatu, SP, Brazil}

\author{Jos\'e Carlos Merino Mombach} 
\affiliation{Centro de Ci\^encias Rurais, Unipampa/S\~ao Gabriel - P\'os-Gradua\c{c}\~ao em F\'isica, Pr\'edio 13, Universidade Federal de Santa Maria, Santa Maria 97105-900, Brazil}
\author{Renata Vieira}
\affiliation{Laborat\'orio de Bioinform\'atica e Biologia Computacional
  Programa Interdisciplinar de Computa\c{c}\~ao Aplicada, Universidade do
  Vale do Rio dos Sinos, 93022-000 S\~ao Leopoldo, RS, Brazil}
\author{Jos\'e Camargo da Silva} 
\affiliation{Laborat\'orio de Bioinform\'atica e Biologia Computacional
  Programa Interdisciplinar de Computa\c{c}\~ao Aplicada, Universidade do
  Vale do Rio dos Sinos, 93022-000 S\~ao Leopoldo, RS, Brazil}
\author{Ney Lemke}
\affiliation{Department of Physics and Biophysics, Institute of Biosciences, S\~ao Paulo State University, UNESP, 18618-000, Botucatu, SP, Brazil}
\author{Marialva Sinigaglia}
\affiliation{Laborat\'orio de Bioinform\'atica e Biologia Computacional
  Programa Interdisciplinar de Computa\c{c}\~ao Aplicada, Universidade do
  Vale do Rio dos Sinos, 93022-000 S\~ao Leopoldo, RS, Brazil}

\date{\today}

\begin{abstract}
  The identification of genes essential for survival is important for
  the understanding of the minimal requirements for cellular life and
  for drug design. As experimental studies with the purpose of
  building a catalog of essential genes for a given organism are
  time-consuming and laborious, a computational approach which could
  predict gene essentiality with high accuracy would be of great value. We
  present here a novel computational approach, called \textit{NTPGE}
  (Network Topology-based Prediction of Gene Essentiality), that
  relies on network topology features of a gene to estimate its
  essentiality. The first step of \textit{NTPGE} is to construct the
  integrated molecular network for a given organism comprising protein
  physical, metabolic and transcriptional regulation interactions. The
  second step consists in training a decision tree-based machine
  learning algorithm on known essential and non-essential genes of the
  organism of interest, considering as learning attributes the network
  topology information for each of these genes. Finally, the decision
  tree classifier generated is applied to the set of genes of this organism
  to estimate essentiality for each gene. We applied the
  \textit{NTPGE} approach for discovering essential genes in
  \textit{Escherichia coli} and then assessed its performance.
  
  \end{abstract}
   %encontrar pacs
  \maketitle
  \section{Introduction}

  Essential genes are genes that are indispensable to support cellular
  life. These genes constitute a minimal set of genes required for a
  living cell. Therefore, the functions encoded by this gene set are
  essential and could be considered as a foundation of life itself
  \cite{kobayashi2003,itaya1995}. The identification of
  genes which are essential for survival is important not only for the
  understanding of the minimal requirements for cellular life, but
  also for practical purposes. For example, since most antibiotics
  target essential cellular processes, essential gene products of
  microbial cells are promising new targets for such drugs
  \cite{jud2000}. The prediction and discovery of
  essential genes has been performed by experimental procedures such
  as single gene knockouts \cite{PubMed_12140549}, RNA
  interference \cite{PubMed_15877598} and conditional knockouts
  \cite{PubMed_14507372}, but each of these techniques require a
  large investment of time and resources and they are not always
  feasible.

  Considering these experimental constraints, a computational or
  \textit{in silico} approach capable of accurately predicting gene
  essentiality would be of great value. Some of such predictors have been
  already developed in which sequence features of genes and proteins
  with or without homology comparison have been
  utilized as parameters for training machine learning classifiers for
  gene essentiality prediction \cite{PubMed_16899653,PubMed_17052348}. 
  In addition, predictors of gene essentiality based on
  network topology features, as the physical interactions of a protein
  \cite{PubMed_11333967} or the number of biochemical species that
  are knocked out from the metabolic network following a gene deletion
  \cite{PubMed_15671116,PubMed_16095595} have also been
  developed.

  The currently available network topology-based methodologies of gene
  essentiality prediction use only one type of network topology
  feature, i.e. protein physical interactions or metabolic
  interactions, for performing such predictions. Actual molecular
  interaction networks, however, are composed by entities that are
  intricately connected with diverse types of interactions, such as
  protein physical, metabolic and transcriptional regulation
  interactions.

  We therefore propose here a novel machine-learning based \textit{in
    silico} approach, called \textit{NTPGE} (Network Topology-based
  Prediction of Gene Essentiality), that relies on multiple
  topological network features of a given gene to estimate its
  essentiality. For the generation of the decision tree classifier,
  NTPGE employs the following network topological features as learning
  attributes: number of physical interactions for the corresponding
  encoded protein, number of target genes transcriptionally regulated
  by the corresponding encoded transcription factor, number of
  transcription factors that regulate it, number of enzymes that use
  metabolites produced by the corresponding encoded enzyme as
  reactants and number of enzymes that produce metabolites used as
  reactants by the corresponding encoded enzyme. To assess the
  performance of the \textit{NTPGE} approach, we used it for the
  discovery of essential genes in the bacterium \textit{Escherichia
    coli}, a model organism whose most of genes have already been
  characterized experimentally as essential or non-essential.

  \section{Construction of the IMN of \textit{E. coli}}
  \label{sec:imn ecoli}

  As \textit{NTPGE} relies on topological features of molecular
  network, the first step was to construct the \textit{Escherichia
    coli} integrated molecular network (IMN) comprising protein
  physical, metabolic and transcriptional regulation interactions. For
  this purpose, we used MONET (MOlecular NETwork) ontology, a tool
  developed by our group that facilitates the construction of IMNs of
  organisms via integration of information from metabolic pathways,
  protein-protein interaction networks and transcriptional regulation
  interactions through a model able to minimize data redundancy and
  inconsistency \cite{PubMed_16755509}. As previously described, two genes of a
  given organism, $g_1$ and $g_2$, coding for proteins $p_1$ and $p_2$
  are linked if:
  \begin{itemize}
  \item $p_1$ and $p_2$ interact physically,
  \item $g_1$ regulates the transcription of gene $g_2$,
  \item or one metabolite generated by a reaction catalyzed by $p_1$
    is consumed in a reaction catalyzed by $p_2$ (we may exclude from
    this analysis the most used compounds such as ATP, NAD, H2O,
    etc.).
  \end{itemize}

  The data sources present in MONET ontology used for the construction
  of the \textit{E. coli} IMN were KEGG (Kyoto Encyclopedia of Genes
  and Genomes \cite{PubMed_16381885} for metabolic interactions, RegulonDB
  \cite{PubMed_16381895} for transcriptional regulation interactions, and Butland
  et al \cite{PubMed_15690043} for protein physical interactions.

  Using MONET, we constructed two directed IMNs of \textit{E. coli},
  $G_a$ and $G_p$. $G_a$ contained all possible interactions among
  genes with 1,998 genes and 51,642 interactions. $G_p$ was similar to $G_a$, except that the
  connections through the ten most frequently used compounds on the
  metabolism were deleted producing a network with 1,987 genes and 21,338 interactions, since connections via
  these common compounds is not likely to be important for the
  determination of gene essentiality due to their promiscuity.

  \section{Brief analysis of the \textit{Escherichia coli} IMNs}
  \label{sec:netanalysis}

  Prior to use the \textit{Escherichia coli} IMNs  
  for the validation of the \textit{NTPGE} approach, we present here a brief analysis of the
  most common network measures, i.e. degree distribution and
  clustering coefficient, of these IMNs. The degree distribution,
  $P(k)$, gives the probability that a selected node has exactly $k$
  links. $P(k)$ is obtained by counting the number of nodes $N(k)$
  with $k$ = 1, 2,... links and dividing by the total number of nodes
  $N$. The clustering coefficient, $C_i$, gives the density of
  triangles we can construct in the network having the node $i$ as a
  vertex. The clusterization coefficient is defined as:
  \begin{equation}
    \label{eq:clusterization}
    C_i=\frac{2n_i}{k_i(k_i-1)}\;,
  \end{equation}
  where $n_i$ is the number of links connecting the $k_i$ neighbors of
  the node $i$. The average clustering coefficient $C$ is the
  clustering coefficient for the whole network and characterizes the
  overall tendency of nodes to form clusters or groups.

\begin{figure}[H]
    \begin{center}
      \includegraphics[scale=0.4, angle=0]{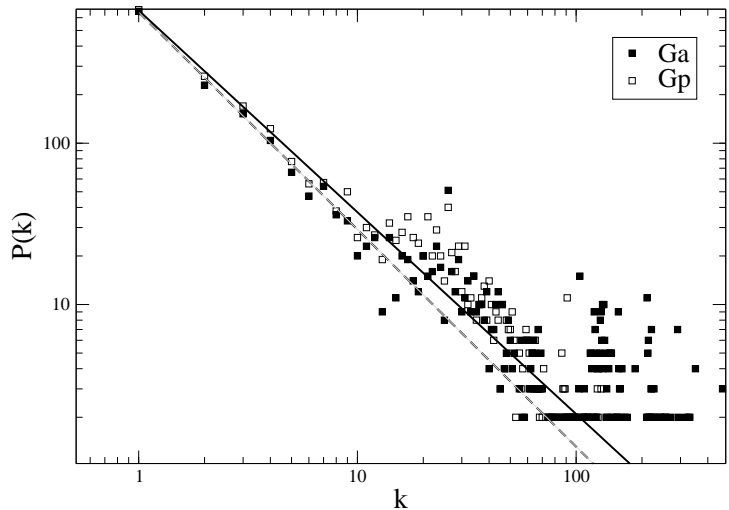}
    \end{center}
    \caption{Histogram of the degree distribution for $G_a$ and $G_p$
      used in this work. Both $G_a$ (solid line) and $G_p$ (dashed line) are well described by a power law
      function $P(k)=Ak^{-\gamma}$ that characterizes them as
      scale-free networks}.
    \label{fig:his-con}
  \end{figure}

  In Figure \ref{fig:his-con} we show the histogram of degree
  distribution for $G_a$ and $G_p$. The curves are well approximated
  by a power law function, $P(k)=Ak^{-\gamma}$ for both IMNs,
  suggesting that $G_a$ and $G_p$ are scale-free networks.

\begin{figure}[H]
    \begin{center}
      \includegraphics*[scale=0.4, angle=0]{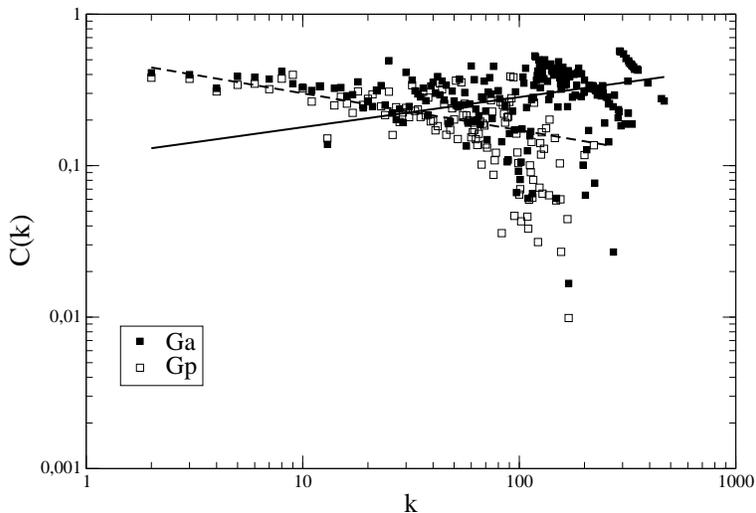}
    \end{center}
    \caption{The dependence of the average clusterization coefficient
      $C$ on the connectivity $k$. The best-fit regression line 
      for $G_a$ (solid line) has a regression 
      slope of $-0.03$ with a confidence interval of
  $[-0.08,0.01]$, while the best-fit regression line for $G_p$ (dashed line) 
  has a regression slope of $0.28$ with a confidence interval of
  $[0.22,0.33]$. The results show that $G_a$ is a 
  non-hierarchical scale-free network, whereas $G_p$ 
  is an hierarchical scale-free network.}
    \label{fig:cxk}
  \end{figure}

  We also analyzed the dependence of the average clusterization
  coefficient, $C$, on the connectivity $k$, defined as $C(k)$. For a
  traditional scale-free network, we expect $C(k)$ not to depend on
  $k$, while for hierarchical networks we expect $C(k)\sim
  k^{-\alpha}$. Figure \ref{fig:cxk} shows the $C(k)$ for $G_a$ and
  $G_p$. These results point to a $C(k)$
  not dependent on $k$ for $G_a$ and a $C(k)$ dependent on $k$ for
  $G_p$, thus indicating that $G_a$ is a non-hierarchical IMN and
  $G_p$ is an hierarchical IMN. This shift from a non-hierarchical
  topology for $G_a$ to an hierarchical topology for $G_p$ seems to be
  caused by the deletion of the connections through the ten most
  frequently used compounds in the metabolism on the construction of
  $G_p$. Such compounds induce a strongly connected IMN due to their
  promiscuity.

  \section{Description of the \textit{NTPGE} approach}
  \label{sec:gene essential}
 
  The \textit{NTPGE} approach was performed using WEKA (\emph{Waikato
    Environment for Knowledge Analysis}) system
  \cite{WEKA}. WEKA is a collection of machine learning
  algorithms for data mining tasks. It also provides means for data
  pre-processing, classification, regression, clustering, association
  rules, and visualization \cite{WEKA}. Among these
  algorithms, we used the J48 \cite{WEKA}, which is the Weka's
implementation of the well known C4.5 \cite{C45} that uses the greedy
  technique to induce decision trees for classification. A
  decision-tree model is built by analyzing training data, which is then used to classify unseen data.

  We trained the J48 algorithm on four different training
  configurations ($t_1$, $t_2$, $t_3$ and $t_4$). In all
  configurations, the training data was a set of known essential and
  non-essential genes of \textit{Escherichia coli} taken from the PEC
  database (\emph{Profiling of Escherichia coli chromosome,
    http://www.shigen.nig.ac.jp/ecoli/pec/}). The PEC
  database has been compiled experimental information on
  \textit{Escherichia coli} strains from research reports and deletion
  mutation studies prior to 1998, including gene essentiality for cell
  growth. Based on these reports about gene essentiality for cell
  growth, the \textit{E, coli} genes are classified in
  essential, non-essential and unknown.
http://www.shigen.nig.ac.jp/ecoli/pec
  In all training configurations, for a given gene, the learning
  attributes used were as follows:

  \begin{itemize}

  \item number of physical interactions for the corresponding encoded protein;
  \item number of target genes transcriptionally regulated by the corresponding encoded transcription factor ({\tt regulation\_out});
  \item number of transcription factors that regulate it; ({\tt regulation\_in});
  \item number of enzymes that use metabolites produced by the corresponding encoded enzyme as reactants ({\tt metabolism\_out});
  \item number of enzymes that produce metabolites used as reactants by the corresponding encoded enzyme ({\tt metabolism\_in});

  \end{itemize}

  In $t_1$ and $t_2$, the above mentioned attributes were extracted
  from $G_a$, whereas these same attributes were extracted from $G_p$
  in $t_3$ and $t_4$. Moreover, the attribute
  \textit{damage}, which was not originally present in $G_a$ and
  $G_p$, was included in $t_2$ and $t_4$, The damage $d$ is defined as the number
  of metabolites whose production was prevented by the deletion of the
  enzyme. For a given enzyme, its damage $d$ has been shown to be
  strongly correlated to its essentiality.\cite{PubMed_14693817}.

  The J48 algorithm was trained with the
  parameters presented in Table I. As it has been known that
  data imbalance is one of the causes that degrade the performance of
  machine learning algorithms \cite{kang}, we replicated
  the data related to the essential genes in order to correct data
  imbalance as the number of non-essential genes is much larger than
  the number of essential genes.

  \section{Performance of the \textit{NTPGE} approach and related
    discussion}
  \label{sec:resdisc}

  The performance of the \textit{NTPGE} approach was evaluated by
  testing the classifiers created by the J48 algorithm, as described
  above, on the training data itself. The selection of the best
  training configuration to be considered as default by the
  \textit{NTPGE} approach was performed based on the
  \textit{F-measure} of the corresponding generated classifier. The
  \textit{F-measure} provides an harmonic mean of precision and recall
  and is defined as:
  \begin{equation}
    \label{eq:f-measure}
    F=\frac{2\times{\mbox{precision}}\times{\mbox{recall}}}{{\mbox{precision}}+{\mbox{recall}}}\;,
    \end{equation}
  
Precision (the percentage of correctly classified instances) and
  recall (the percentage of positive labeled instances that were
  classified as such) were calculated from the confusion matrices of
  the classifiers obtained from the training configurations $t_1$,
  $t_2$, $t_3$ and $t_4$ (Tables II) and are shown
  on Table III. Table III also shows the \textit{F-measure} as well as
  the features of the training configurations, as the number of
  instances (genes plus metabolites) and presence or absence of the
  learning attribute damage $d$ on training.

  According to Table III, the best training configuration was $t_1$
  (all genes and metabolites with the attribute damage). Its
  corresponding generated classifier had a \textit{F-measure} of
  83.4\% for essential genes and 79.7\% for non-essential genes. In
  fact, all generated classifiers yielded similar results, suggesting
  that the presence or absence of the ten most used compounds in
  metabolism or the presence or absence of the attribute damage $d$
  did not affect the classification of genes as essential
  or non-essential by the \textit{NTPGE} approach. Therefore, any
  training configuration could be selected as default by
  \textit{NTPGE}.

  \begin{table}[htb]
\caption{\label{tab:parametros} Parameters used to run the J48 algorithm on training data.}    
\begin{center}\begin{tabular}{lr}
        \hline
        {\sf Parameter} & {\sf Value}\\
        \hline       
        {\sf binarySplit} & {\sf False}\\
        
        {\sf confidenceFactor} & {\sf 0.25}\\
        
        {\sf debug} & {\sf False}\\
        
        {\sf minNumObj} & {\sf 100}\\
        
        {\sf numFolds} & {\sf 3}\\
        
        {\sf reduceErrorPruning} & {\sf False}\\
        
        {\sf saveInstanceData} & {\sf False}\\
        
        {\sf seed} & {\sf 1}\\
        
        {\sf subtreeRaising} & {\sf True}\\
        
        {\sf unpruned} & {\sf False}\\
        
        {\sf useLaplace} & {\sf False}\\
        \hline
      \end{tabular}
    \end{center}
   \end{table}
  
\begin{table}[htb]
\caption{\label{tab:r1}Confusion matrices of the classifiers generated from $t_1$, $t_2$, $t_3$ and $t_4$}    

\begin{center}\begin{tabular}{ccc}
\hline \hline
{\sf} & {\sf $t_1$} & {\sf} \\        
\hline \hline
        \multicolumn{2}{c}{\sf Predicted} \\ 
\cline{1-2}
{\sf Non-essential} & {\sf Essential} & {\sf Actual} \\
        \hline
        {\sf 1,392} & {\sf 397} & {\sf Non-essential}\\
        {\sf 310} & {\sf 1,780} & {\sf Essential}\footnotemark[1] \\
\hline \hline
{\sf} & {\sf $t_2$} & {\sf} \\ \hline \hline        
        \multicolumn{2}{c}{\sf Predicted} \\ 
\cline{1-2}
{\sf Non-essential} & {\sf Essential} & {\sf Actual} \\
        \hline
        {\sf 1,348} & {\sf 405} & {\sf Non-essential} \\
        \hline
        {\sf 313} & {\sf 1,777} & {\sf Essential}\footnotemark[1] \\
        \hline \hline
{\sf} & {\sf $t_3$} & {\sf} \\ \hline \hline
\multicolumn{2}{c}{\sf Predicted} \\ 
\cline{1-2}
{\sf Non-essential} & {\sf Essential} & {\sf Actual} \\       
        \hline
        {\sf 1,346} & {\sf 432} & {\sf Non-essential}\\
        \hline
        {\sf 298} & {\sf 1,792} & {\sf Essential}\footnotemark[1] \\
        \hline \hline
{\sf} & {\sf $t_4$} & {\sf} \\ \hline \hline
\multicolumn{2}{c}{\sf Predicted} \\
\cline{1-2}
{\sf Non-essential} & {\sf Essential} & {\sf Actual} \\
        \hline
        {\sf 1,348} & {\sf 430} & {\sf Non-essential}\\
        \hline
        {\sf 300} & {\sf 1,790} & {\sf Essential}\footnotemark[1] \\
        \hline      
\end{tabular}
\footnotetext[1]{The number of essential genes were replicated to avoid data imbalance. Actually, the number of essential genes is 209}

\end{center}
    
  \end{table}

\begin{table}[htb]
\caption{\label{tab:result} {\sf Features of the training configurations and performance measures of their corresponding generated classifiers}.}    
\begin{center}\begin{tabular}{ccccc}
        \hline
        {\sf Features and Performance Measures} & \multicolumn{4}{c}{\sf Training configurations}\\
	\cline{2-5} 
	& {\sf$t_1$} & {\sf$t_2$} & {\sf$t_3$} & {\sf$t_4$}\\ 	
        \hline
        
        {\sf Number of Genes}\footnotemark[1] & {\sf 3,879} & {\sf 3,879} &
        {\sf 3,868} & {\sf 3,868}\\
        {\sf Damage $d$} & {\sf no} & {\sf yes} &
        {\sf no} & {\sf yes}\\
          
        {\sf Correctly Predicted Genes (\%)} & {\sf 81.8} & {\sf 81.5} &
        {\sf 81.1} & {\sf 81.1}\\
          
        {\sf Incorrectly Predicted Genes (\%)} & {\sf 18.2} & {\sf 18.5} &
        {\sf 18.9} & {\sf 18.9}\\
          
        {\sf F-measure (N) (\%)} & {\sf\textbf{79.7}} & {\sf 79.4} &
        {\sf 78.7} & {\sf 78.7}\\
          
        {\sf F-measure (E) (\%)} & {\sf\textbf{83.4}} & {\sf 83.2} &
        {\sf 83.1} & {\sf 83.1}\\
          
        {\sf Recall (N) (\%)} & {\sf 77.8} & {\sf 77.4} &
        {\sf 75.7} & {\sf 75.8}\\
          
        {\sf Recall (E) (\%)} & {\sf 85.2} & {\sf 85.0} &
        {\sf 85.7} & {\sf 85.6}\\
          
        {\sf Precision (N) (\%)} & {\sf 81.8} & {\sf 81.6 } &
        {\sf 81.9} & {\sf 81.8}\\
          
        {\sf Precision (E) (\%)} & {\sf 81.8} & {\sf 81.4} &
        {\sf 80.6} & {\sf 80.6}\\
        \hline
      \end{tabular}
\footnotetext[1]{The number of essential genes were replicated to avoid data imbalance; number of non-essential genes remained unchanged. Actually, the number of essential genes is 209 and non-essential genes is 1,789 for $G_a$ and the number of essential genes is 209 and non-essential genes is 1,778 for $G_p$}
\end{center}
    
  \end{table}

\begin{figure}[H]
    \begin{center}
      \includegraphics[width=0.8\columnwidth, keepaspectratio]
      {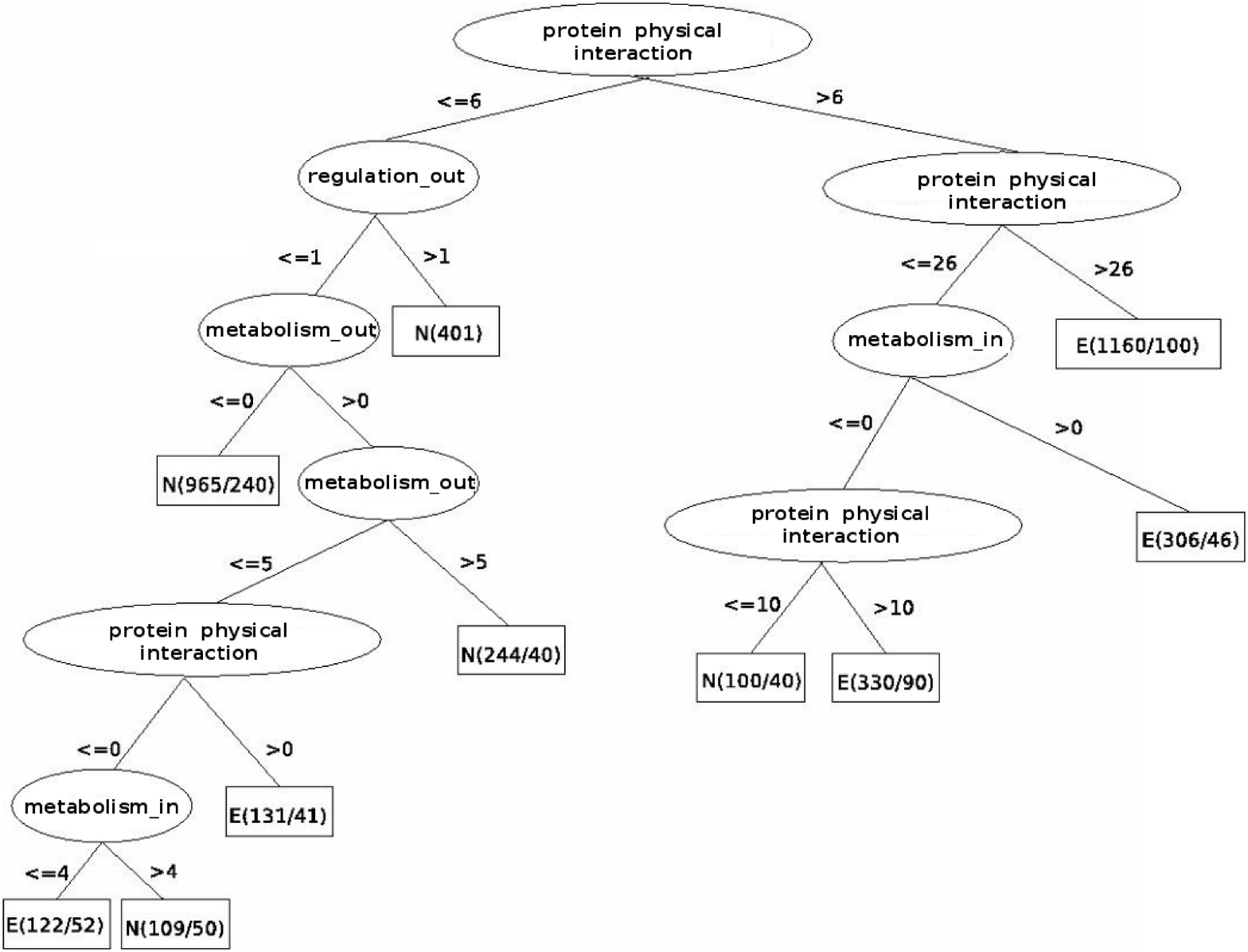}
    \end{center}
    \caption{Decision tree generated from $t_1$ with a
      \textit{F-measure} of 83.4\% for essential genes (E) and 79.7\% for non-essential genes (N). The \textit{($x$/$y$)} inside rectangles denotes the number of correctly classified genes ($x$) and the number of incorrectly classified genes ($y$).}
    \label{fig:sem10_replicado}
  \end{figure}

  Figure \ref{fig:sem10_replicado} shows the set of rules of the
  decision tree generated from $t_1$. The top node of the tree
  corresponds to the attribute protein physical interaction. This
  means that the classification tree algorithm concluded that the main factor to define essentiality in \textit{E. coli} was the protein
  physical interaction. In fact, the degree of a protein has been
  documented in the literature as being indicative of essentiality in
  various organisms \cite{PubMed_11333967,PubMed_16281187,PubMed_15231746}. In our approach, a combination of
  intermediate number of protein physical interactions with at least
  one interaction of the type {\tt metabolism\_in}, i.e. number of
  enzymes that produce metabolites used as reactants by the
  corresponding encoded enzyme, was also indicative of essentiality. Transcriptional regulation interactions seems not to be a good predictor for gene essentiality, since genes with at least one interaction of
  the type {\tt regulation\_out}, i.e. number of target genes
  transcriptionally regulated by the corresponding encoded
  transcription factor, were classified as non-essential. Moreover, the attribute ({\tt regulation\_in}, i.e. the number of
  transcription factors that regulate a given gene, was not even
  included in the decision tree. These results regarding gene essentiality and transcriptional�l regulation are not surprising, since transcription factors are usually not essential under the conditions in which the knockout experiments for determining gene essentiality are performed (PEC database, http://www.shigen.nig.ac.jp/ecoli/pec/)

  \section{Concluding remarks}
  \label{sec:conclusion}

  We proposed here a novel machine learning-based computational
  approach, called \textit{NTPGE} (Network Topology-based Prediction
  of Gene Essentiality), that relies on network topology features of a
  gene to estimate its essentiality. Distinct from previous network
  topology-based gene essentiality predictors, \textit{NTPGE} employs
  multiple topological network features of a given gene to estimate
  its essentiality, namely physical interactions for the corresponding encoded protein, number
  of target genes transcriptionally regulated by the corresponding
  encoded transcription factor, number of transcription factors that
  regulate it, number of enzymes that use metabolites produced by the
  corresponding encoded enzyme as reactants and number of enzymes that
  produce metabolites used as reactants by the corresponding encoded
  enzyme.

  We verified the performance of \textit{NTPGE} by applying it for the
  discovery of essential genes in the bacterium \textit{Escherichia
    coli}, a model organism whose most of genes have already been
  characterized experimentally as essential or non-essential. Among
  the interactions considered as learning attributes, \textit{NTPGE}
  relied mostly in protein physical and metabolic interactions for
  gene essentiality prediction. In addition, the presence or absence
  of the ten most used compounds in metabolism or the presence or
  absence of the attribute damage $d$ did not likely influence the
  classification of genes as essential or non-essential by
  \textit{NTPGE}. This can be concluded because the
  \textit{F-measure} values of all generated decision trees were
  similar. Anyway, the best classifier was generated from $t_1$ (all
  genes and metabolites with the attribute damage) with a
  \textit{F-measure} of 83.4\% for essential genes and 79.7\% for
  non-essential genes.

  In conclusion, the \textit{NTPGE} seems to be a reliable method of
  gene essentiality discovery that may be applied to the gene set of
  other organisms. However, \textit{NTPGE} is limited to organisms
  whose corresponding IMN has already been constructed. The
  construction of the IMN of a given organism involves the gathering
  of experimentally determined data that are not always available,
  particularly for a newly sequenced organism. To overcome this
  limitation, future developments would be the integration of
  \textit{NTPGE} with sequence-based methods of IMN construction, thus
  creating a purely \textit{in silico} network topology
  information-based methodology of gene essentiality discovery.

\section*{ACKNOWLEDGEMENTS}
We would like to thank CNPq (research grants 474278/2006-9 and
506414/2004-3), FAPESP (research grant 2007/02827-9) and FAPERGS
(05600005-BRD) for supporting this work. 
We would also like to thank HP Brazil R\&D for the collaboration.

\bibliographystyle{prsty}
%%\bibliography{/home/denimaru/unesp/Manuscripts}
%\bibliography{fisica.bib}

\begin{thebibliography}{10}

\bibitem{kobayashi2003}
K. K., S. Ehrlich, A. Albertini, G. Amati, K. Andersen, M. Arnaud, K. Asai, S.
  Ashikaga, S. Aymerich, P. Bessieres, and et~al, Proc. Natl Acad. Sci. USA
  {\bf 100},  4678  (2003).

\bibitem{itaya1995}
I. M., FEBS Lett. {\bf 362},  257  (1995).

\bibitem{jud2000}
J. N. and J. Mekalanos, Nat. Biotechnol. {\bf 18},  740  (2000).

\bibitem{PubMed_12140549}
G. Giaever, A.~M. Chu, L. Ni, C. Connelly, L. Riles, S. V{\'e}ronneau, S. Dow,
  A. Lucau-Danila, K. Anderson, B. Andr{\'e}, A.~P. Arkin, A. Astromoff, M.
  El-Bakkoury, R. Bangham, R. Benito, S. Brachat, S. Campanaro, M. Curtiss, K.
  Davis, A. Deutschbauer, K.-D. Entian, P. Flaherty, F. Foury, D.~J. Garfinkel,
  M. Gerstein, D. Gotte, U. G{\"u}ldener, J.~H. Hegemann, S. Hempel, Z. Herman,
  D.~F. Jaramillo, D.~E. Kelly, S.~L. Kelly, P. K{\"o}tter, D. LaBonte, D.~C.
  Lamb, N. Lan, H. Liang, H. Liao, L. Liu, C. Luo, M. Lussier, R. Mao, P.
  Menard, S.~L. Ooi, J.~L. Revuelta, C.~J. Roberts, M. Rose, P. Ross-Macdonald,
  B. Scherens, G. Schimmack, B. Shafer, D.~D. Shoemaker, S. Sookhai-Mahadeo,
  R.~K. Storms, J.~N. Strathern, G. Valle, M. Voet, G. Volckaert, C. yun Wang,
  T.~R. Ward, J. Wilhelmy, E.~A. Winzeler, Y. Yang, G. Yen, E. Youngman, K. Yu,
  H. Bussey, J.~D. Boeke, M. Snyder, P. Philippsen, R.~W. Davis, and M.
  Johnston, Nature {\bf 418},  387  (2002).

\bibitem{PubMed_15877598}
L.~M. Cullen and G.~M. Arndt, Immunol Cell Biol {\bf 83},  217  (2005).

\bibitem{PubMed_14507372}
T. Roemer, B. Jiang, J. Davison, T. Ketela, K. Veillette, A. Breton, F. Tandia,
  A. Linteau, S. Sillaots, C. Marta, N. Martel, S. Veronneau, S. Lemieux, S.
  Kauffman, J. Becker, R. Storms, C. Boone, and H. Bussey, Mol Microbiol {\bf
  50},  167  (2003).

\bibitem{PubMed_16899653}
M. Seringhaus, A. Paccanaro, A. Borneman, M. Snyder, and M. Gerstein, Genome
  Res {\bf 16},  1126  (2006).

\bibitem{PubMed_17052348}
A.~M. Gustafson, E.~S. Snitkin, S.~C.~J. Parker, C. DeLisi, and S. Kasif, BMC
  Genomics {\bf 7},  265  (2006).

\bibitem{PubMed_11333967}
H. Jeong, S.~P. Mason, A.~L. Barab{\'a}si, and Z.~N. Oltvai, Nature {\bf 411},
  41  (2001).

\bibitem{PubMed_15671116}
M. Imieli{\'n}ski, C. Belta, A. Hal{\'a}sz, and H. Rubin, Bioinformatics {\bf
  21},  2008  (2005).

\bibitem{PubMed_16095595}
M.~C. Palumbo, A. Colosimo, A. Giuliani, and L. Farina, FEBS Lett {\bf 579},
  4642  (2005).

\bibitem{PubMed_16755509}
J.~P.~M. da~Silva, N. Lemke, J.~C. Mombach, J.~G.~C. de~Souza, M. Sinigaglia,
  and R. Vieira, Genet Mol Res {\bf 5},  182  (2006).

\bibitem{PubMed_16381885}
M. Kanehisa, S. Goto, M. Hattori, K.~F. Aoki-Kinoshita, M. Itoh, S. Kawashima,
  T. Katayama, M. Araki, and M. Hirakawa, Nucleic Acids Res {\bf 34},  D354
  (2006).

\bibitem{PubMed_16381895}
H. Salgado, S. Gama-Castro, M. Peralta-Gil, E. D{\'i}az-Peredo, F.
  S{\'a}nchez-Solano, A. Santos-Zavaleta, I. Mart{\'i}nez-Flores, V.
  Jim{\'e}nez-Jacinto, C. Bonavides-Mart{\'i}nez, J. Segura-Salazar, A.
  Mart{\'i}nez-Antonio, and J. Collado-Vides, Nucleic Acids Res {\bf 34},  D394
   (2006).

\bibitem{PubMed_15690043}
G. Butland, J.~M. Peregr{\'i}n-Alvarez, J. Li, W. Yang, X. Yang, V. Canadien,
  A. Starostine, D. Richards, B. Beattie, N. Krogan, M. Davey, J. Parkinson, J.
  Greenblatt, and A. Emili, Nature {\bf 433},  531  (2005).

\bibitem{WEKA}
I.~H. Witten and E. Frank, {\em Data Mining: Pratical Machine Learning Tools
  and Techniques with Java Implementations} (Morgan Kaufmann, San Francisco,
  2000).

\bibitem{C45}
J.~R. Quinlan, {\em C4.5: programs for machine learning} (Morgan Kaufmann, San
  Francisco, 1993).

\bibitem{PubMed_14693817}
N. Lemke, F. Her{\'e}dia, C.~K. Barcellos, A.~N.~D. Reis, and J.~C.~M. Mombach,
  Bioinformatics {\bf 20},  115  (2004).

\bibitem{kang}
P. Kang and S. Cho, Lecture Notes in Computer Science {\bf 4232},  837  (2006).

\bibitem{PubMed_16281187}
E. Estrada, Proteomics {\bf 6},  35  (2006).

\bibitem{PubMed_15231746}
S. Wuchty, Genome Res {\bf 14},  1310  (2004).

\end{thebibliography}

%\begin{thebibliography}{10}

%\end{thebibliography}

\end{document}